\documentclass[12pt]{article}
\usepackage{amsmath,amssymb,geometry,physics,bm}
\title{Topological-Insulator and Spintronic Boundary Electrodynamics for MRI RF Coils: A Theoretical Framework for Loss, Noise, and Reciprocity}

\author{
Zoran Milosavljevic\\
{\footnotesize Independent Researcher, Johns Hopkins University}
}
\date{}

\begin{document}
\maketitle

\begin{abstract}
MRI radiofrequency (RF) coils are ultimately limited by conductor loss, thermal noise, and reciprocity constraints associated with conventional metallic boundary conditions. These limitations become more severe at higher static fields, where operating frequencies increase and current distributions are governed by surface impedance and electromagnetic coupling in the near field. In this work we develop a theoretical framework that incorporates topological-insulator (TI) surface transport and spintronic interface physics into RF coil electrodynamics. Starting from the Dirac surface Hamiltonian and linear-response (Kubo/Drude) transport, we derive an effective complex surface impedance for TI-coated conductors and establish modified boundary conditions for tangential fields in the presence of spin--momentum locking and spin--charge coupling. We then analyze time-reversal-symmetry-breaking TI/ferromagnet interfaces, where an anomalous Hall surface conductivity produces antisymmetric admittance and enables nonreciprocal RF response. Finally, we connect these results to MRI metrics including coil quality factor, thermal noise, and receive sensitivity through reciprocity-based formulations. The framework identifies parameter regimes in which topological and spintronic surface transport could reduce RF dissipation, modify noise mechanisms, and enable coil-level nonreciprocity without conventional ferrites.
\end{abstract}

\section{Introduction}

RF coils in magnetic resonance imaging (MRI) generate and detect oscillatory magnetic fields at the Larmor frequency $\omega_0 = \gamma B_0$, where $B_0$ is the static field \cite{Hoult1976,Hoult2000}. The coil and its matching network determine transmit efficiency ($B_1^+$ per unit power), receive sensitivity, and ultimately signal-to-noise ratio (SNR). At higher $B_0$ the frequency increases and conductor loss, surface current nonuniformity, and electromagnetic coupling become more pronounced; these effects are commonly described using surface impedance models and reciprocity-based receive formulations \cite{Hoult1976,Roemer1990}.

Conventional coil conductors are normal metals (typically copper), whose RF losses are governed by diffusive transport and the classical skin effect. In the surface-impedance approximation, power dissipation scales with a surface resistance $R_s(\omega)\propto \sqrt{\omega/\sigma}$, where $\sigma$ is bulk conductivity \cite{Pozar}. Thermal noise is linked to dissipation by fluctuation--dissipation principles, implying that any reduction in effective surface resistance can directly reduce coil-generated noise \cite{Hoult1976,Hoult2000}.

Topological insulators (TIs) provide a qualitatively different boundary transport mechanism: insulating bulk with conducting surface states described by massless (or weakly massive) Dirac fermions with spin--momentum locking \cite{HasanKane2010,QiZhang2011,Ando2013}. In the absence of time-reversal symmetry breaking, elastic backscattering is suppressed, enabling large transport mean free paths under suitable conditions. In addition, TI surfaces naturally host spin--charge coupling effects such as the Edelstein effect (current-induced spin accumulation) and inverse Edelstein effect (spin-to-charge conversion) \cite{Edelstein1990,Manchon2015}. When time-reversal symmetry is broken by magnetic proximity or doping, TI surfaces can acquire a Hall response and magnetoelectric terms that modify electromagnetic boundary conditions and can yield nonreciprocal behavior \cite{QiHughesZhang2008,TseMacDonald2010}.

The purpose of this paper is \emph{not} to review TI physics, but to develop a \emph{coil-relevant electrodynamic theory} in which TI and spintronic surface transport enter as modified boundary conditions and effective surface impedance/admittance tensors. We (i) establish baseline RF coil loss/noise relations using surface impedance, (ii) derive an effective RF surface conductivity for TI surfaces using linear response (Drude/Kubo), (iii) translate this into boundary conditions for coil conductors, (iv) analyze Hall-type surface conductivity and resulting nonreciprocity in TI/ferromagnet interfaces, and (v) connect the theory to MRI coil figures of merit using reciprocity formulations \cite{Hoult1976,Roemer1990}.

\section{RF Coil Electrodynamics: Loss, Quality Factor, and Noise}

\subsection{Surface-impedance description of conductor loss}

For an RF conductor whose fields vary slowly along the surface compared with the skin depth, conductor loss may be expressed in terms of the tangential magnetic field $\bm{H}_\parallel$ at the surface using a surface impedance $Z_s(\omega)$ \cite{Pozar}:
\begin{equation}
\bm{E}_\parallel = Z_s(\omega)\, \bm{J}_s,
\qquad
\bm{J}_s = \hat{\bm{n}} \times \bm{H}_\parallel,
\label{eq:surfimp}
\end{equation}
where $\bm{J}_s$ is surface current density and $\hat{\bm{n}}$ is the outward normal.

Time-averaged conductor power dissipation is
\begin{equation}
P_{\mathrm{cond}} = \frac{1}{2}\int_S \Re\{Z_s(\omega)\}\, |\bm{J}_s|^2\, dS
= \frac{1}{2}\int_S \Re\{Z_s(\omega)\}\, |\hat{\bm{n}}\times\bm{H}_\parallel|^2\, dS.
\label{eq:Pcond}
\end{equation}
For a good normal metal, $Z_s \approx (1+i)\sqrt{\omega\mu/(2\sigma)}$ \cite{Pozar}, giving
\begin{equation}
R_s^{\mathrm{metal}}(\omega)=\Re\{Z_s\} = \sqrt{\frac{\omega\mu}{2\sigma}}.
\label{eq:Rsmetal}
\end{equation}
Equation \eqref{eq:Pcond} is the entry point for incorporating \emph{alternative surface transport} into RF loss models: once a physically motivated $Z_s(\omega)$ is specified, the loss follows from the same electromagnetic formalism.

\subsection{Coil quality factor and partition of losses}

In resonant operation, the unloaded quality factor is $Q_0=\omega_0 W/P_{\mathrm{loss}}$, where $W$ is stored energy and $P_{\mathrm{loss}}$ total dissipated power. In MRI coils, loss is often partitioned into conductor loss and sample loss \cite{Hoult1976,Hoult2000}:
\begin{equation}
\frac{1}{Q_0} = \frac{1}{Q_{\mathrm{cond}}} + \frac{1}{Q_{\mathrm{sample}}} + \frac{1}{Q_{\mathrm{other}}}.
\end{equation}
At low frequencies or small coils, conductor loss can dominate; at higher frequencies and typical human imaging loads, sample loss often dominates, but conductor loss remains important for small coils, preamplifier-decoupled arrays, and high-$Q$ structures \cite{Roemer1990,Hoult2000}. Materials or boundary conditions that reduce $R_s$ can increase $Q_{\mathrm{cond}}$ and reduce coil-generated noise contributions.

\subsection{Thermal noise and fluctuation--dissipation}

Coil thermal noise is tied to dissipation. In simple circuit models, the voltage noise spectral density of a resistor is $4k_BTR$ (per unit bandwidth). In distributed RF structures, the fluctuation--dissipation theorem implies that the same dissipative mechanisms described by $\Re\{Z_s\}$ generate Johnson noise fields/currents \cite{Hoult1976,Hoult2000}. Thus, if a modified boundary condition reduces dissipative surface resistance without introducing new loss channels, it can reduce the coil noise contribution and improve SNR in coil-noise-dominated regimes.

\section{Receive Sensitivity and Reciprocity in MRI}

\subsection{Reciprocity-based receive formulation}

A central result in MRI coil theory is that receive sensitivity can be expressed via electromagnetic reciprocity \cite{Hoult1976}. In one common form, the induced receive voltage is proportional to the transverse magnetic field the coil would produce per unit current at the point of the spins. For modern multi-channel arrays, this reciprocity picture underlies sensitivity and noise-covariance modeling \cite{Roemer1990}. Reciprocity-based derivations assume linear, time-invariant, \emph{reciprocal} media and boundary conditions.

If coil materials introduce \emph{nonreciprocal} response (e.g., via antisymmetric conductivity tensors), then reciprocity relations must be generalized. This is not necessarily detrimental, but it changes how transmit and receive sensitivities relate, and it can alter noise correlations. The framework developed below makes explicit where nonreciprocity enters: through boundary admittance tensors with antisymmetric components.

\subsection{Why nonreciprocity is attractive for RF coils}

MRI coil arrays require decoupling between elements and isolation between coil and preamplifier to avoid noise coupling and instability \cite{Roemer1990}. Conventional nonreciprocal components (circulators/isolators) rely on ferrites and magnetic bias fields, which are challenging to integrate at the coil level in MRI environments. If a TI/ferromagnet boundary can provide effective nonreciprocity in a thin-film geometry \cite{TseMacDonald2010}, it could enable \emph{coil-integrated nonreciprocal behavior} without bulky ferrites (subject to MRI compatibility constraints). The analysis in Sec.\ \ref{sec:TRbreak} formalizes this possibility at the boundary-condition level.

\section{Topological-Insulator Surface Transport Model}

\subsection{Dirac surface Hamiltonian and helical states}

The simplest model for a 3D TI surface is a single Dirac cone:
\begin{equation}
H_0 = \hbar v_F (\hat{\bm{z}}\times \bm{\sigma})\cdot \bm{k},
\label{eq:Dirac}
\end{equation}
with eigenstates exhibiting spin--momentum locking \cite{HasanKane2010,QiZhang2011}. In time-reversal-invariant conditions, backscattering from nonmagnetic disorder is suppressed by spin structure, often increasing the effective transport time compared with conventional surfaces.

For RF electrodynamics, we require the linear response of surface current to an in-plane electric field at frequencies $\omega$ relevant to MRI (typically $\sim$ tens to hundreds of MHz). In the regime $\hbar \omega \ll E_F$ and for moderate temperatures, intraband response dominates and a Drude-like form provides a useful starting point \cite{Mahan}.

\subsection{Surface conductivity in Drude/Kubo form}

A minimal phenomenological model for TI surface conductivity is
\begin{equation}
\sigma_s(\omega) = \frac{\sigma_{s0}}{1-i\omega\tau},
\qquad
\sigma_{s0} = e^2 D(E_F) v_F^2 \tau,
\label{eq:Drude}
\end{equation}
where $\tau$ is an effective momentum relaxation time and $D(E_F)$ is the surface density of states at the Fermi energy. For a Dirac dispersion $E=\hbar v_F k$, the 2D density of states scales as $D(E_F)\propto |E_F|/(2\pi \hbar^2 v_F^2)$, giving $\sigma_{s0}\propto (e^2/h)\, (E_F\tau/\hbar)$ up to order-unity factors (a standard Dirac-fermion scaling) \cite{Mahan}.

Equation \eqref{eq:Drude} may be viewed as a low-frequency limit of Kubo linear response. The important point for RF coils is that $\Re\{\sigma_s\}$ determines dissipation and depends directly on $\tau$. Mechanisms that increase $\tau$ (e.g., reduced backscattering) can reduce dissipative impedance.

\subsection{From surface conductivity to surface impedance}

For a purely 2D conducting sheet with surface conductivity $\sigma_s(\omega)$, a useful effective boundary relation is
\begin{equation}
\bm{J}_s = \sigma_s(\omega)\, \bm{E}_\parallel,
\qquad
\Rightarrow
\bm{E}_\parallel = Z_s^{\mathrm{TI}}(\omega)\, \bm{J}_s,
\qquad
Z_s^{\mathrm{TI}}(\omega)=\frac{1}{\sigma_s(\omega)}.
\label{eq:Zsheet}
\end{equation}
Thus,
\begin{equation}
Z_s^{\mathrm{TI}}(\omega) = \frac{1-i\omega\tau}{\sigma_{s0}}
= \underbrace{\frac{1}{\sigma_{s0}}}_{R_s^{\mathrm{TI}}}
- i\underbrace{\frac{\omega\tau}{\sigma_{s0}}}_{X_s^{\mathrm{TI}}}.
\label{eq:ZTI}
\end{equation}
The dissipative part is $R_s^{\mathrm{TI}}=1/\sigma_{s0}$, while $X_s^{\mathrm{TI}}$ is reactive. In contrast to normal metals, where $R_s\propto \sqrt{\omega}$ due to skin effect \cite{Pozar}, a strictly 2D Drude sheet has a frequency-independent resistive part in this regime and a reactive part linear in $\omega$.

Real TI-coated conductors are not ideal 2D sheets: there may be bulk conduction, finite thickness, and coupling to underlying metal. Nonetheless, Eq.\ \eqref{eq:ZTI} provides a first-principles path to replacing the classical $Z_s$ with a TI-derived $Z_s$ in Eq.\ \eqref{eq:Pcond}, thereby linking condensed-matter transport to RF dissipation.

\section{TI-Coated Conductors: Composite Boundary Conditions}

\subsection{Motivation for composite (metal + TI) boundaries}

An MRI coil requires a low-impedance current path. A plausible architecture is a conventional conductor that is coated or interfaced with a TI layer such that a significant fraction of RF current flows in (or is influenced by) TI surface channels. The relevant electrodynamics is then a \emph{composite boundary problem}: a bulk conductor with its own surface impedance in parallel/series with a surface sheet admittance.

A common effective model treats the TI surface as a sheet admittance $Y_s=\sigma_s$ placed at the boundary of a conductor characterized by a classical surface impedance $Z_{\mathrm{bulk}}$ \cite{Pozar}. For tangential fields, an approximate parallel combination gives an effective surface impedance
\begin{equation}
Z_{\mathrm{eff}}(\omega)\approx \left(\frac{1}{Z_{\mathrm{bulk}}(\omega)} + \sigma_s(\omega)\right)^{-1}.
\label{eq:Zeff}
\end{equation}
If $\sigma_s$ is sufficiently large (and/or if $Z_{\mathrm{bulk}}$ is sufficiently lossy), the sheet channel can appreciably reduce $\Re\{Z_{\mathrm{eff}}\}$.

\subsection{Implications for current distribution}

The surface current density obeys $\bm{J}_s=\hat{\bm{n}}\times \bm{H}_\parallel$. A reduced $Z_{\mathrm{eff}}$ changes how the RF current distributes over conductor surfaces for fixed applied voltages and matching constraints. In loop coils, for example, reduced dissipation can increase $Q_{\mathrm{cond}}$ and reduce coil noise. In arrays, it can also influence coupling and decoupling networks via changed conductor loss and effective resistive loading \cite{Roemer1990}.

\subsection{Order-of-magnitude scaling comparison}

Although material-specific parameters are required for quantitative prediction, the scaling difference is notable:

\begin{itemize}
\item Metal skin-effect: $R_s^{\mathrm{metal}}(\omega)\propto \sqrt{\omega}$ \cite{Pozar}.
\item TI sheet (Drude): $R_s^{\mathrm{TI}}\sim 1/\sigma_{s0}\propto 1/\tau$ (weak frequency dependence in low-$\omega$ limit) \cite{Mahan}.
\end{itemize}

Thus, as $\omega$ increases, a TI-dominated boundary could exhibit a slower increase (or no increase) in resistive loss in the regime where the sheet model is valid, potentially favoring higher-field MRI.

\section{Spin--Charge Coupling and Spintronic Boundary Terms}

\subsection{Edelstein effect and coupled response}

Spin--momentum locking implies that a charge current produces a nonequilibrium spin density (Edelstein effect), and conversely, injected spin can generate a charge current (inverse Edelstein effect) \cite{Edelstein1990,Manchon2015}. A minimal linear coupled-response form is
\begin{equation}
\bm{J}_c = \sigma_s \bm{E}_\parallel + \lambda\, (\hat{\bm{z}}\times \bm{s}),
\qquad
\bm{s} = \chi\, (\hat{\bm{z}}\times \bm{E}_\parallel) - \Gamma \bm{s},
\label{eq:spincharge}
\end{equation}
where $\bm{s}$ is surface spin density (or spin accumulation), $\chi$ parameterizes current-induced spin polarization, and $\Gamma$ captures relaxation. In the frequency domain, spin relaxation introduces dispersion, effectively modifying the reactive and resistive parts of the boundary response.

The coil-relevant point is that the boundary condition is no longer purely Ohmic: the effective surface admittance can acquire additional frequency dependence from spin dynamics, particularly if the TI is coupled to a magnetic layer whose ferromagnetic resonance (FMR) lies near the RF frequency range (a possibility in some engineered systems).

\subsection{Effect on RF impedance and dissipation}

Eliminating $\bm{s}$ from Eq.\ \eqref{eq:spincharge} yields an effective conductivity
\begin{equation}
\sigma_{\mathrm{eff}}(\omega) = \sigma_s(\omega) + \Delta\sigma(\omega),
\end{equation}
where $\Delta\sigma$ encodes spin-charge conversion and spin relaxation. In general, $\Re\{\Delta\sigma\}$ contributes additional dissipation channels (spin pumping, magnetic damping), while $\Im\{\Delta\sigma\}$ modifies stored energy. The sign and magnitude depend on microscopic coupling and relaxation rates \cite{Manchon2015}.

For MRI coils, where noise is critical, the central theoretical question becomes: under what parameter regimes does $\Delta\sigma$ reduce net dissipation (e.g., by redistributing current away from lossy bulk paths) versus increase dissipation (e.g., by introducing magnetic damping losses)? The framework here makes those questions explicit via boundary admittance terms.

\section{Time-Reversal Symmetry Breaking, Hall Conductivity, and Nonreciprocity}
\label{sec:TRbreak}

\subsection{Gapped surface states and conductivity tensor}

If time-reversal symmetry is broken on the TI surface (magnetic proximity or doping), the Dirac Hamiltonian acquires a mass term:
\begin{equation}
H = H_0 + \Delta\, \sigma_z,
\end{equation}
opening a gap $2|\Delta|$ \cite{QiZhang2011}. In such systems, the surface conductivity becomes tensorial:
\begin{equation}
\bm{\sigma} =
\begin{pmatrix}
\sigma_{xx} & \sigma_{xy} \\
-\sigma_{xy} & \sigma_{xx}
\end{pmatrix}.
\label{eq:sigtensor}
\end{equation}
The antisymmetric Hall term $\sigma_{xy}$ is the key ingredient for nonreciprocal boundary response. Related electromagnetic consequences for TI surfaces and magnetoelectric response have been analyzed in the context of topological magnetoelectric effects and Faraday/Kerr rotation \cite{QiHughesZhang2008,TseMacDonald2010}.

\subsection{Boundary conditions with Hall sheet conductivity}

For a conducting sheet with tensor conductivity, the surface current is
\begin{equation}
\bm{J}_s = \bm{\sigma}\, \bm{E}_\parallel.
\end{equation}
Equivalently, the boundary condition relating $\bm{E}_\parallel$ and $\bm{J}_s$ involves an \emph{impedance tensor} $\bm{Z}_s=\bm{\sigma}^{-1}$. The antisymmetric part implies that an applied tangential field generates a surface current rotated relative to $\bm{E}_\parallel$, producing effective gyration. This is precisely the structure that produces nonreciprocity in RF/microwave systems (conceptually analogous to magneto-optic gyrotropy), but here it arises from a thin-film boundary rather than bulk ferrites \cite{TseMacDonald2010}.

\subsection{Nonreciprocal response and reciprocity theorem implications}

Electromagnetic reciprocity relies on symmetry of the underlying constitutive relations. An antisymmetric conductivity tensor violates the conditions of standard Lorentz reciprocity. In MRI contexts, this means transmit and receive sensitivities may no longer be related by the simplest reciprocity statement \cite{Hoult1976}. Instead, one must use generalized reciprocity formulations for nonreciprocal media, in which the adjoint problem involves transposed constitutive tensors.

Practically, this suggests two outcomes:
\begin{itemize}
\item Nonreciprocity could be exploited for element isolation/decoupling, potentially improving array stability.
\item The relationship between coil-generated $B_1^+$ and receive sensitivity must be interpreted carefully, especially if nonreciprocal effects are strong.
\end{itemize}
The present work does not attempt a full MRI reconstruction formalism for nonreciprocal coils, but it identifies the electrodynamic origin of reciprocity modification and provides the correct boundary-condition language to build such a theory.

\section{Connection to MRI Figures of Merit}

\subsection{Quality factor and conductor-noise reduction}

Given an effective surface impedance $Z_{\mathrm{eff}}(\omega)$ (e.g., Eq.\ \eqref{eq:Zeff} or tensor generalizations), the conductor loss follows from Eq.\ \eqref{eq:Pcond} with $R_s\to \Re\{Z_{\mathrm{eff}}\}$. Coil quality factor and coil noise then follow from standard RF relations \cite{Pozar} and MRI noise arguments \cite{Hoult2000}. The framework implies that improved $\tau$ (suppressed backscattering) increases $\sigma_{s0}$ and decreases $R_s^{\mathrm{TI}}$, which can increase $Q_{\mathrm{cond}}$.

\subsection{Sample noise vs coil noise regimes}

In many clinical MRI applications, sample noise dominates \cite{Hoult2000}. However, for small coils (surface coils, preclinical imaging, local receive arrays), coil losses remain significant \cite{Roemer1990}. Additionally, in high-$Q$ receive systems (cryogenic coils, or specialized resonant structures), conductor noise can be a limiting factor. The present theory is most impactful in regimes where conductor loss materially contributes to total noise or where coil-level nonreciprocity offers system-level benefits (decoupling/isolation).

\subsection{High-field relevance}

As $\omega$ increases with $B_0$, the metal $R_s^{\mathrm{metal}}\propto \sqrt{\omega}$ grows \cite{Pozar}. In contrast, the low-$\omega$ TI sheet resistance in Eq.\ \eqref{eq:ZTI} is not intrinsically $\sqrt{\omega}$-scaling. This motivates the hypothesis that TI-influenced boundaries could become increasingly favorable at higher fields, provided the TI surface channel remains the dominant RF current path and bulk conduction/disorder are controlled \cite{Ando2013}.

\section{Limitations, Assumptions, and Practical Theoretical Extensions}

The present framework makes several simplifying assumptions:

\begin{itemize}
\item \textbf{Surface-dominated transport:} Bulk conduction in the TI is neglected or treated as subdominant; in real materials, bulk carriers can be significant \cite{Ando2013}.
\item \textbf{Linear response:} We assume linear conductivity (Kubo/Drude). Nonlinear spintronic effects and strong-field regimes are not treated.
\item \textbf{Local boundary condition:} The sheet model assumes local relation between $\bm{J}_s$ and $\bm{E}_\parallel$; nonlocal effects (finite mean free path compared with feature size) may modify results \cite{Mahan}.
\item \textbf{Magnetic compatibility:} TI/ferromagnet interfaces may introduce stray fields and magnetic noise; a complete MRI feasibility analysis would require incorporating these noise sources.
\end{itemize}

Despite these limitations, the boundary-condition formulation is valuable: it provides the correct theoretical knobs ($\tau$, $E_F$, $\sigma_{xy}$, spin relaxation rates) and the correct electrodynamic objects (scalar/tensor $Z_s$) needed for quantitative simulations and experimental design.

\section{Conclusion}

We developed a theoretical framework linking topological-insulator and spintronic surface transport to RF coil electrodynamics relevant to MRI. By expressing TI surface transport through an effective complex surface conductivity and translating it into (possibly tensorial) surface impedance boundary conditions, we connected condensed-matter surface-state physics to RF dissipation, coil quality factor, and thermal noise. We further analyzed time-reversal-symmetry breaking TI/ferromagnet interfaces, where Hall-type surface conductivity introduces antisymmetric boundary admittance and enables nonreciprocal RF response, implying modified reciprocity relations central to MRI receive theory. The framework identifies regimes in which topological and spintronic boundaries could reduce RF losses or enable coil-level isolation, motivating future quantitative modeling and experimental validation.

\section*{Acknowledgments}
The author thanks colleagues and reviewers in the condensed-matter and MRI RF engineering communities for helpful discussions that informed the framing of this work. The author also acknowledges the developers and maintainers of the open-source \TeX\ ecosystem and related scientific computing tools that support modern technical writing and analysis.

\end{document}